\pdfoutput=1
\documentclass[11pt]{article}

\usepackage[final]{acl}

\usepackage{fancyhdr}
\fancypagestyle{firstpage}{
  \fancyhf{}
    \fancyhead[L]{\normalsize Accepted in ACL 2025}
}

\usepackage{times}
\usepackage{latexsym}

\usepackage[T1]{fontenc}
\usepackage[utf8]{inputenc}

\usepackage{microtype}

\usepackage{inconsolata}

\usepackage{graphicx}
\usepackage{booktabs}
\usepackage{adjustbox}
\usepackage{ulem}
\usepackage{amsmath}
\usepackage{amssymb}  % Also includes \mathbb

\definecolor{darkgreen}{RGB}{0, 100, 0} % RGB values for dark green
\usepackage{multirow} % For \multirow
\title{Hard Negative Mining for Domain-Specific Retrieval in Enterprise Systems}

\author{
 \textbf{Hansa Meghwani\textsuperscript{*}},
 \textbf{Amit Agarwal\textsuperscript{*}},
 \textbf{Priyaranjan Pattnayak}, \\%\textsuperscript{2}}, \\
 \textbf{Hitesh Laxmichand Patel}, %\textsuperscript{2}},
 \textbf{Srikant Panda} %\textsuperscript{1}}
\\
\\
Oracle AI
\\
 \small{
   \textbf{Correspondence:} \href{mailto:email@domain}{hansa.meghwani@oracle.com; amit.h.agarwal@oracle.com}
 }
 \thanks{\textsuperscript{}The authors contributed equally to this work.}
}

\begin{document}
\thispagestyle{firstpage}
\pagestyle{firstpage}
\maketitle

\begin{abstract}
% Enterprise search systems require accurate retrieval of domain-specific information, yet semantic mismatches and overlapping terminologies often lead traditional models to retrieve irrelevant content, negatively impacting applications such as knowledge management, customer support, and retrieval-augmented generation agents. 
Enterprise search systems often struggle to retrieve accurate, domain-specific information due to semantic mismatches and overlapping terminologies. These issues can degrade the performance of downstream applications such as knowledge management, customer support, and retrieval-augmented generation agents. To address this challenge, we propose a scalable hard-negative mining framework tailored specifically for domain-specific enterprise data. Our approach dynamically selects semantically challenging but contextually irrelevant documents to enhance deployed re-ranking models.

Our method integrates diverse embedding models, performs dimensionality reduction, and uniquely selects hard negatives, ensuring computational efficiency and semantic precision. Evaluation on our proprietary enterprise corpus (cloud services domain) demonstrates substantial improvements of 15\% in MRR@3 and 19\% in MRR@10 compared to state-of-the-art baselines and other negative sampling techniques. Further validation on public domain-specific datasets (FiQA, Climate Fever, TechQA) confirms our method's generalizability and readiness for real-world applications.%industry deployment.
\end{abstract}

\section{Introduction}

Accurate retrieval of domain-specific information significantly impacts critical enterprise processes, such as knowledge management, customer support, and Retrieval Augmented Generation (RAG) Agents. However, achieving precise retrieval remains challenging due to semantic mismatches, overlapping terminologies, and ambiguous abbreviations common in specialized fields like finance, and cloud computing. Traditional lexical retrieval techniques, such as BM25~\cite{Robertson1994}, struggle due to vocabulary mismatches, leading to irrelevant results and poor user experience.

Recent dense retrieval approaches leveraging pre-trained language models, like BERT-based encoders~\cite{Karpukhin2020,Xiong2020,Guu2020}, mitigate lexical limitations by capturing semantic relevance. Nevertheless, their performance heavily relies on the negative samples—documents incorrectly retrieved due to semantic similarity but lacking contextual relevance. Models trained with negative sampling methods (e.g., random sampling, BM25-based static sampling, or dynamic methods like ANCE~\cite{Xiong2020}, STAR~\cite{Zhan2021}) either lack sufficient semantic discrimination or incur high computational costs, thus limiting scalability and practical enterprise deployment. For instance, given a query such as \textit{"Steps to deploy a MySQL database on Cloud Infrastructure,"} most negative sampling techniques  select documents discussing non-MySQL database deployments. Conversely, our method strategically selects a hard negative discussing MySQL deployment on-premises, which despite semantic overlap, is contextually distinct and thus poses a stronger training challenge for the retrieval and re-ranking models.

Our proposed framework addresses these by introducing a novel semantic selection criterion explicitly designed to curate high-quality hard negatives. % Unlike prior methods relying solely on static lexical matching (BM25) or computationally intensive fully dynamic methods (ANCE, STAR), 
By uniquely formulating two semantic conditions that effectively select negatives that closely resemble query semantics but remain contextually irrelevant, significantly minimizing false negatives encountered by existing techniques. 
\noindent The main contributions of this paper are:

\begin{enumerate}
\item A negative mining framework for dynamically selecting semantically challenging hard negatives, leveraging diverse embedding models and semantic filtering criteria to significantly improve re-ranking models in domain-specific retrieval scenarios.

\item Comprehensive evaluations demonstrating consistent and significant improvements across both proprietary and publicly available datasets, verifying our method's impact and broad applicability across domain-specific usecases.

\item In-depth analysis, of critical challenges in handling both short and long-form enterprise documents, laying a clear foundation for targeted future improvements. 
\end{enumerate}

Our work directly enhances the semantic discrimination capabilities of re-ranking models, resulting in \textbf{15\% improvement in MRR@3} and \textbf{19\% improvement in MRR@10} on our in-house cloud-services domain dataset. Further evaluations on public domain-specific benchmarks (FiQA, Climate Fever, TechQA) confirm generalizability and tangible improvements of our proposed negative mining framework.

\section{Related Work}

\subsection{Hard Negatives in Retrieval Models}
The role of hard negatives in training dense retrieval models has been widely studied. Static negatives, such as BM25~\cite{Robertson1994}, provide lexical similarity but fail to capture semantic relevance, often leading to overfitting~\cite{Qu2020}. Dynamic negatives, introduced in ANCE~\cite{Xiong2020} and STAR~\cite{Zhan2021}, adapt during training to provide more challenging contrasts but require significant computational resources due to periodic re-indexing. Our framework addresses these limitations by dynamically identifying semantically challenging negatives using clustering and dimensionality reduction, ensuring scalability and adaptability.

Further studies have explored advanced methods for negative sampling in cross-encoder models~\cite{meghwani2024enhancingretrievalperformanceensemble}. Localized Contrastive Estimation (LCE)~\cite{Guo2023} integrates hard negatives into cross-encoder training, improving the reranking performance when negatives align with the output of the retriever. Similarly, \cite{Pradeep2022} demonstrated the importance of hard negatives even when models undergo advanced pretraining techniques, such as condenser~\cite{Gao2021}. Our work builds on these efforts by offering a scalable approach, which can be applied to any domain-heavy enterprise data.%enterprise which has domain-heavy knowledge base.

\subsection{Negative Sampling Strategies}
Effective negative sampling significantly affects the performance of the retrieval model by challenging the model to differentiate between relevant and irrelevant examples. Common strategies include:
\begin{itemize}
    \item \textbf{Random Negatives:} Efficient but lacking semantic contrast, leading to suboptimal performance~\cite{Karpukhin2020}.
    \item \textbf{BM25 Negatives:} Leverage lexical similarity, but often introduce biases, particularly in semantically rich domains~\cite{Robertson1994}.
    \item \textbf{In-Batch Negatives:} Computationally efficient but limited to local semantic contrasts, often underperforming in dense retrieval tasks~\cite{Xiong2020}.
\end{itemize}

Our framework complements these approaches by dynamically generating negatives that balance semantic similarity and contextual irrelevance, avoiding the pitfalls of static or random methods.

\subsection{Domain-Specific Retrieval Challenges}
Enterprise retrieval systems face unique challenges, such as ambiguous terminology, overlapping concepts, and private datasets~\cite{meghwani2024enhancingretrievalperformanceensemble}. General-purpose methods such as BM25 or dense retrieval models~\cite{Qu2020} fail to capture domain-specific complexities effectively. Our approach addresses these gaps by curating hard negatives that align with enterprise-specific semantics, improving retrieval precision and robustness for proprietary datasets. 

\noindent We further discuss negative sampling techniques in Appendix~\ref{sec:appendix_related_works}.

% \textbf{Extended Discussion:} For a detailed comparison of static and dynamic strategies, as well as synthetic data generation methods, see Appendix~\ref{sec:appendix_related_works}.

\begin{figure*}[th!]
    \centering
    \includegraphics[width=\textwidth]{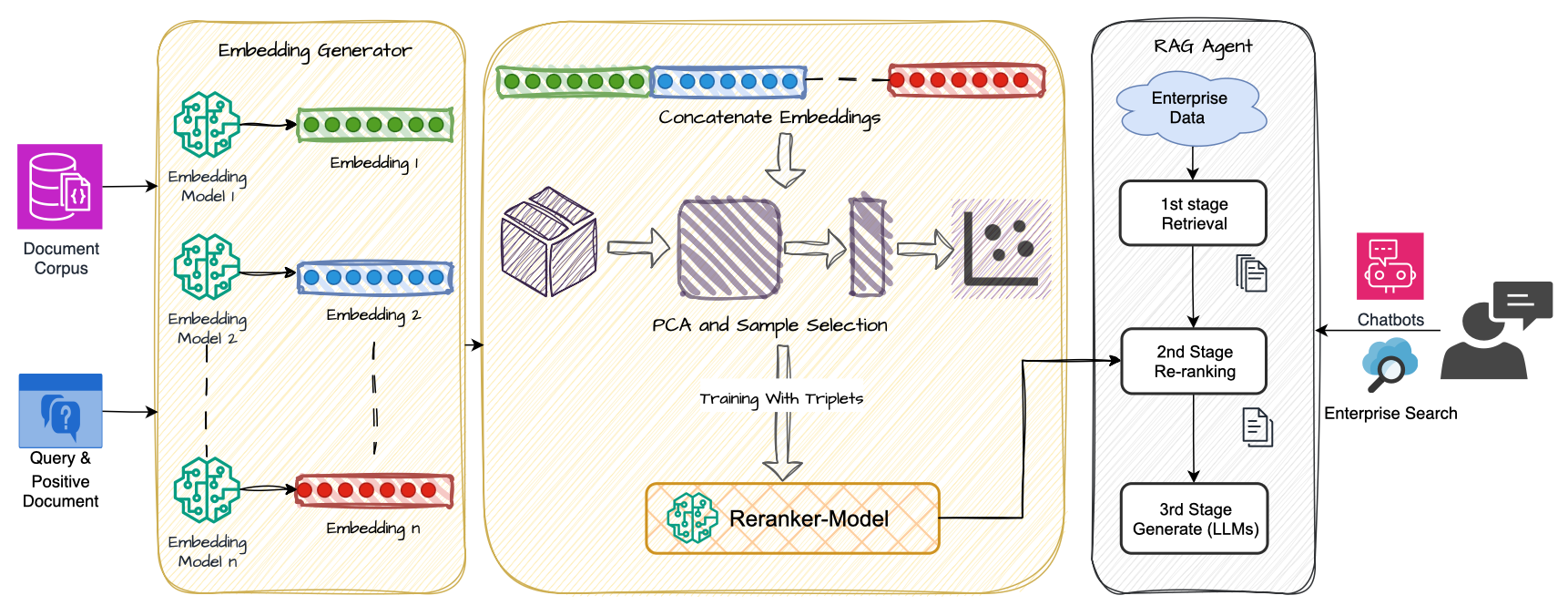} % Placeholder for pipeline diagram
    \caption{Overview of the methodology pipeline for training reranker models, including embedding generation, PCA-based dimensionality reduction and hard negative selection for fine-tuning.}
    \label{fig:pipeline_placeholder}
    \vspace{-1em}
\end{figure*}

\section{Methodology}

To effectively train and finetune reranker models for domain-specific retrieval, it is essential to systematically handle technical ambiguities stemming from specialized terminologies, overlapping concepts, and abbreviations prevalent within enterprise domains.

We propose a structured, modular framework that integrates diverse embedding models, dimensionality reduction, and a novel semantic criterion for hard-negative selection. Figure \ref{fig:pipeline_placeholder} illustrates the high-level pipeline, components and their interactions. The re-ranking models fine-tuned using the hard negatives generated by our framework are directly deployed in downstream applications, such as RAG, significantly improving the resolution of customer queries through enhanced retrieval.

Our approach begins by encoding queries and documents into semantically rich vector representations using an ensemble of state-of-the-art bi-encoder embedding models. These embeddings are strategically selected based on multilingual support, embedding quality, training data diversity, context length handling, and performance (details provided in Appendix %\ref{appendix:embedding_models})
\ref{sec:extended_methodoly}. To manage embedding dimensionality and improve computational efficiency, Principal Component Analysis (PCA) \cite{mackiewicz1993principal} is utilized to project the concatenated embeddings onto a lower-dimensional space, maintaining 95\% of the original variance. 

We then define two semantic conditions (Eq.~\ref{eq:q_distance} and Eq.~\ref{eq:pd_distance}) to dynamically select high-quality hard negatives, addressing semantic similarity challenges and minimizing false negatives. Together, these two equations ensure that the selected hard negative is not only close to the query (Eq.~\ref{eq:q_distance}) but also contextually distinct from the true positive, minimizing the risk of selecting topic duplicates or noisy positives  (Eq.~\ref{eq:pd_distance}). For example, a query about deploying MySQL on Oracle Cloud, PD is a guide on that topic, and D is a doc about MySQL on-premise — semantically close to Q, but distant from PD.

Below we detail each methodological component, emphasizing their contributions to enhancing retrieval precision in domain-specific or enterprise retrieval tasks.

\begin{table}[h!]
\centering
\begin{tabular}{lcccc}
\toprule
                     & \textbf{Total} & \textbf{Train} & \textbf{Test} \\ \midrule
\(<Q, PD>\)             & 5250            & 1000            & 4250           \\ 
\bottomrule
\end{tabular}
\caption{Dataset distribution of queries (Q) and positive documents (PD).}
\vspace{-1em}
\label{tab:dataset_stats}
\end{table}

\subsection{Dataset Statistics}

Our experiments leverage a proprietary corpus containing 36,871 unannotated documents sourced from over 30 enterprise cloud services. Additionally, we prepared 5250 annotated query-positive document pairs (\(<Q, PD>\)) for training and testing. Notably, we adopted a non-standard train-test split (as summarized in Table \ref{tab:dataset_stats}), allocating four times more data to testing than training to rigorously evaluate model robustness against varying training data volumes (additional analyses in Appendix~\ref{appendix:training_datasize}). To further validate generalizability, we conduct evaluations on publicly available domain-specific benchmarks: FiQA (finance) \cite{fiqa_dataset}, Climate Fever (climate science) \cite{diggelmann2021climatefeverdatasetverificationrealworld}, and TechQA (technology) \cite{castelli2019techqadataset}. 
Detailed dataset statistics are provided in Appendix~\ref{appendix:data_stats}.

\subsection{Embedding Generation}

Embeddings for queries, positive documents, and the corpus are computed via six diverse, high-performance bi-encoder models \(E_1, E_2, \dots, E_6\), each selected strategically for capturing complementary semantic perspectives:
\begin{equation}
\mathbf{E}_k(x) \in \mathbb{R}^{d_k}
\label{eq:1}
\end{equation}

where $d_k$ is the embedding dimension of the $k_{th}$ model for textual input $x$. Concatenation of these embeddings yields a comprehensive representation:
\begin{equation}
\mathbf{X}_{\text{concat}} = [\mathbf{e}_1(x); \mathbf{e}_2(x); \dots; \mathbf{e}_6(x)] %\in \mathbb{R}^{\sum_{k=1}^6 d_k}
\label{eq:2}
\end{equation}

where $\mathbf{X}_{\text{concat}} \in \mathbb{R}^{\sum_{k=1}^6 d_k}$ represents the concatenated embedding for the input $x$.% obtained by stacking individual embeddings $\mathbf{e}_k(x)$ horizontally.

\subsection{Dimensionality Reduction}

To alleviate the computational overhead arising from high-dimensional concatenated embeddings, we apply PCA to reduce dimensionality while preserving semantic richness:
\begin{equation}
\mathbf{X}_{\text{PCA}} = \mathbf{X}_{\text{concat}} \mathbf{P},
\label{eq:3}
\end{equation}

where \(\mathbf{P}\) represents the PCA projection matrix. We specifically select PCA due to its computational efficiency, and scalability, essential given our large enterprise corpus and high-dimensional embedding space. While we empirically evaluated nonlinear dimensionality reduction methods such as UMAP \cite{mcinnes2020umapuniformmanifoldapproximation} and t-SNE \cite{van2008visualizing}, they offered negligible performance improvements over PCA but incurred substantially higher computational costs, making them impractical for deployment at scale in enterprise systems.

\subsection{Hard Negative Selection Criteria}

We propose two semantic criteria to identify high-quality hard negatives. PCA-reduced embeddings \(\mathbf{X}_{\text{PCA}}\) are organized around each query \(Q\). For each query-positive document pair \((Q, PD)\), candidate documents \(D\) from the corpus are evaluated via cosine distances:
\begin{equation}
d(Q, PD), \quad d(Q, D), \quad d(PD, D)
\label{eq:distance}
\end{equation}

A document \(D\) is selected as a hard negative only if it satisfies both criteria:
\begin{align}
d(Q, D) &< d(Q, PD) \label{eq:q_distance} \\
d(Q, D) &< d(PD, D) \label{eq:pd_distance}
\end{align}

% Equation \eqref{eq:q_distance} ensures the candidate is semantically closer to the query than the positive document, indicating potential confusion for the reranker. Equation \eqref{eq:pd_distance} confirms the candidate document is sufficiently distinct from the positive document (another potential positive document), preventing selection of false negatives that could degrade model training.

Equation~\eqref{eq:q_distance} ensures that the candidate negative document is semantically closer to the query than the actual positive document, making it a challenging negative example that potentially confuses the reranking model. Equation~\eqref{eq:pd_distance}, ensures that the selected hard negative is not just query-confusing but also sufficiently dissimilar from the actual positive (avoiding near-duplicates or false negatives).

The candidate document \(D_{HN}\) with minimal \(d(Q, D)\) satisfying these conditions is chosen as the primary hard negative. Additional hard negatives can similarly be selected based on semantic proximity rankings.

\begin{figure}[h]
    \centering
    \includegraphics[width=1.1\columnwidth]{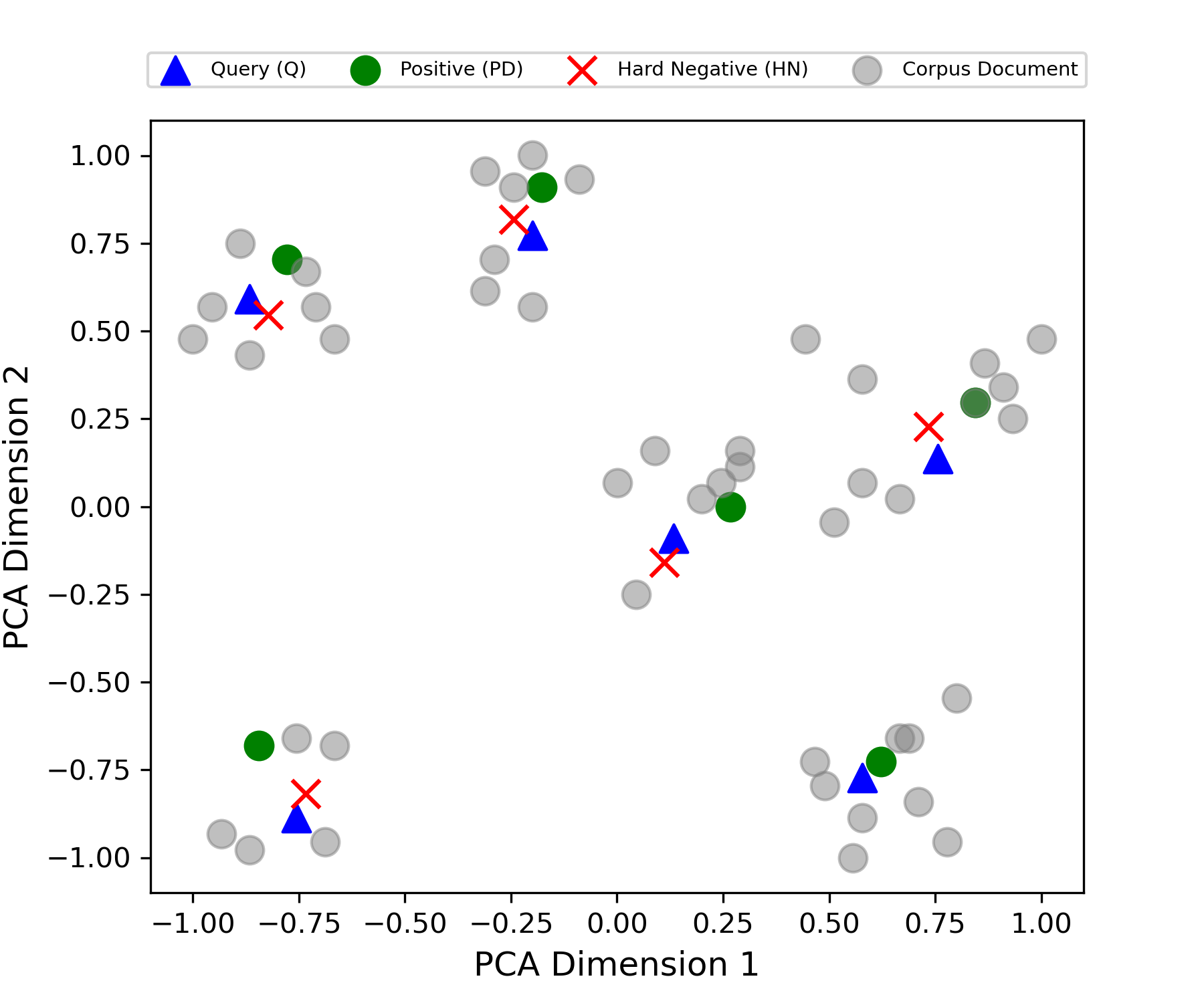} % Replace with your figure file
    \caption{Hard negative selection on the first two PCA components (78\% variance). $Q$ act as centroids, $PD$ guide selection of hard negatives; which are chosen based on semantic proximity.}
    \label{fig:cluster_analysis}
    \vspace{-1em}
\end{figure}

Figure \ref{fig:cluster_analysis} illustrates an example embedding space, clearly depicting the query \(Q\), positive document \(PD\), and selected hard negative \(D_{HN}\), visualizing the semantic selection criteria. In cases where no documents satisfy these conditions, no hard negatives are selected for that particular query. Further details on our embedding model \& fine-tuning using these hard negatives are provided in Appendix \ref{sec:extended_methodoly}.

\section{Experiments \& Results}

To evaluate the effectiveness of our proposed hard-negative selection framework, we conduct extensive experiments on our internal cloud-specific enterprise dataset, as well as domain-specific open-source benchmarks. We systematically compare our approach against multiple competitive negative sampling methods and perform detailed ablation studies to understand the contribution of individual framework components. Complete details on experimental setups and hyperparameters are provided in Appendix \ref{appenix:setup}.

\subsection{Results \& Discussion}

\begin{table*}[!ht]
    \centering
    \scalebox{0.8}{ 
    \begin{tabular}{lcc|cc|cc|cc}
        \toprule
        \multirow{2}{*}{Re-ranker (Fine-tuned w/)} & \multicolumn{2}{c}{Internal} & \multicolumn{2}{c}{FiQA} & \multicolumn{2}{c}{Climate-FEVER} & \multicolumn{2}{c}{TechQA} \\
        & MRR@3 & MRR@10 & MRR@3 & MRR@10 & MRR@3 & MRR@10 & MRR@3 & MRR@10 \\
        \midrule                                                                                          
        Baseline (No Fine-tuning) & 0.42 & 0.45 & 0.45 & 0.48 & 0.44 & 0.46 & 0.57 & 0.61 \\
        In-batch Negatives  & 0.47 & 0.52 & 0.46 & 0.52 & 0.44 & 0.47 & 0.57 & 0.62 \\
        STAR  & 0.53 & 0.56 & 0.51 & 0.54 & 0.47 & 0.49 & 0.61 & 0.63 \\
        ADORE+STAR  & 0.54 & 0.57 & 0.52 & 0.54 & 0.48 & 0.52 & 0.63 & 0.66 \\
        \textbf{Our Proposed HN} & \textbf{0.57} & \textbf{0.64} & \textbf{0.54} & \textbf{0.56} & \textbf{0.52} & \textbf{0.55} & \textbf{0.65} & \textbf{0.69} \\  
        \bottomrule
    \end{tabular}
    }
    \caption{Comparative performance benchmarking of our in-house reranker across multiple domain-specific datasets. The reranker is fine-tuned (FT) with different negative sampling techniques, highlighting the effectiveness of our proposed hard-negative mining method (HN).}
    \label{tab:dataset_benchmark}
    % \vspace{-1em}
\end{table*}

\begin{table}[!ht]
\centering
\resizebox{\columnwidth}{!}{
\begin{tabular}{lcc}
\toprule
\textbf{Negative Sampling Method} & \textbf{MRR@3} & \textbf{MRR@10} \\ \midrule
Baseline                           & 0.42           & 0.45           \\
FT with Random Neg                 & 0.47           & 0.51           \\
FT with BM25 Neg                   & 0.49           & 0.54           \\
FT with In-batch Neg               & 0.47           & 0.52           \\
FT with BM25+In-batch Neg          & 0.52           & 0.54           \\
FT with STAR                               & 0.53           & 0.56           \\
FT with ADORE+STAR                         & 0.54           & 0.57           \\
FT with our HN                               & \textbf{0.57}  & \textbf{0.64}  \\ \bottomrule
\end{tabular}}
\caption{Comparison of negative sampling methods for fine-tuning(FT) in-house cross-encoder reranker model. The proposed framework achieves 15\% and 19\% improvements in MRR@3 and MRR@10, respectively, over baseline methods.}
\label{tab:negative_sampling}
\vspace{-1em}
\end{table}

\paragraph{Comparative Analysis of Negative Sampling Strategies}
Table~\ref{tab:negative_sampling} presents a detailed comparison of of our negative sampling technique against several established methods, including Random, BM25, In-batch, STAR, and ADORE+STAR. The baseline is defined as the performance of our internal reranker model without any fine-tuning. Our method achieves notable relative improvements of {15\% in MRR@3} and {19\% in MRR@10} over this baseline. The semantic nature of our hard negatives allows the reranker to distinguish contextually irrelevant but semantically similar documents effectively. In contrast, simpler baselines like Random or BM25 negatives suffer due to no semantic consideration, while advanced methods like STAR and ADORE+STAR occasionally miss subtle semantic nuances that our formulated selection criteria address effectively.

% A qualitative review of queries revealed that our method particularly excels in cases where subtle contextual differences exist, highlighting its value in domain-specific and enterprise scenarios requiring nuanced retrieval. (NEED EXAMPLE)% precision. For instance, queries involving ambiguous abbreviations or domain-specific terminology frequently demonstrate clear advantages when our proposed negatives are utilized.

\paragraph{Generalization Across Open-source Models}
To validate the robustness and versatility of our framework, we evaluated various open-source embedding and reranker models (Table~\ref{tab:open_source_encoders}), clearly demonstrating improvements across all models when fine-tuned using our proposed negative sampling compared to ADORE+STAR and baseline (no fine-tuning). Notably, rerankers with multilingual capabilities, such as the BGE-Reranker and Jina Reranker, demonstrated pronounced improvements, likely benefiting from our embedding ensemble's multilingual semantic richness. Similarly, larger models like e5-mistral exhibit significant gains, reflecting their capacity to exploit nuanced semantic differences provided by our negative samples. This analysis underscores the general applicability and model-agnostic benefits of our approach.

\begin{table}[!h]
\centering

\label{tab:reranker-benchmark}
\scalebox{0.65}{
\begin{tabular}{lccc}
\toprule
\textbf{Model} & \textbf{Baseline} & \textbf{ADORE+STAR} & \textbf{Ours} \\
\midrule
Alibaba-NLP \\ (gte-multilingual-reranker-base) & 0.39 & 0.42 & \textbf{0.45} \\
\hline
BGE-Reranker\\ (bge-reranker-large) & 0.44 & 0.47 & \textbf{0.52} \\
\hline
Cohere Embed English Light \\(Cohere-embed-english-light-v3.0) & 0.32 & 0.34 & \textbf{0.38} \\
\hline
Cohere Embed Multilingual \\(Cohere-embed-multilingual-v3.0) & 0.34 & 0.37 & \textbf{0.40} \\
\hline
Cohere Reranker\\ (rerank-multilingual-v2.0) & 0.42 & 0.45 & \textbf{0.49} \\
\hline
IBM Reranker\\ (re2g-reranker-nq) & 0.40 & 0.43 & \textbf{0.46} \\
\hline
Infloat Reranker\\ (e5-mistral-7b-instruct) & 0.35 & 0.38 & \textbf{0.42} \\
\hline
Jina Reranker v2 \\(jina-reranker-v2-base-multilingual) & 0.45 & 0.48 & \textbf{0.53} \\
\hline
MS-MARCO\\ (ms-marco-MiniLM-L-6-v2) & 0.41 & 0.43 & \textbf{0.46} \\
\hline
Nomic AI Embed Text \\(nomic-embed-text-v1.5) & 0.33 & 0.36 & \textbf{0.39} \\
\hline
NVIDIA\\ NV-Embed-v2 & 0.38 & 0.41 & \textbf{0.44} \\
\hline
Salesforce\\ SFR-Embedding-2\_R & 0.37 & 0.40 & \textbf{0.43} \\
\hline
Salesforce\\ SFR-Embedding-Mistral & 0.36 & 0.39 & \textbf{0.42} \\
\hline
T5-Large & 0.41 & 0.44 & \textbf{0.47} \\
\bottomrule
\end{tabular}}
\caption{Performance benchmarking (MRR@3) of reranker and embedding models using the proposed hard negative selection framework, compared with ADORE+STAR and baseline methods.}
\label{tab:open_source_encoders}
\vspace{-2em}
\end{table}

\paragraph{Effectiveness on Domain-specific Public Datasets}
We further tested our method's adaptability across diverse public domain-specific datasets (FiQA, Climate-FEVER, TechQA), as shown in Table~\ref{tab:dataset_benchmark}. Each dataset presents distinct retrieval challenges, ranging from technical jargon in TechQA to complex domain-specific reasoning in Climate-FEVER. Fine-tuning with our generated hard negatives consistently improved retrieval across these varied datasets. FiQA exhibited significant gains, likely due to the semantic differentiation required in finance-specific queries. These results demonstrate that our negative sampling method is not only effective within our internal enterprise corpus but also valuable across diverse, domain-specific public datasets, indicating broad applicability and domain independence.

\begin{table}[th!]
\centering
\scalebox{0.75}{
\begin{tabular}{llcc}
\toprule
\textbf{}                          & \textbf{Model}          & \textbf{MRR@3} & \textbf{MRR@10} \\ \midrule
\multirow{2}{*}{\textbf{Short Documents}} & Baseline             & 0.481          & 0.526           \\
                                   & FT w/ proposed HN         & \textbf{0.61}  & \textbf{0.662}  \\ \midrule
\multirow{2}{*}{\textbf{Long Documents}}  & Baseline             & 0.423          & 0.477           \\
                                   & FT w/ proposed HN         & \textbf{0.475} & \textbf{0.521}  \\ \bottomrule
\end{tabular}}

\caption{Performance comparison of the in-house reranker without fine-tuning (Baseline) versus fine-tuned (FT) with our proposed hard negatives (HN), evaluated separately on short and long documents.}

\label{tab:short_long_docs}
\vspace{-1em}
\end{table}

\paragraph{Performance Analysis on Short vs. Long Documents}
An explicit analysis of short versus long documents (Table~\ref{tab:short_long_docs}) revealed differential performance gains. Short documents (under 1024 tokens) experienced substantial performance improvements (MRR@3 improving from 0.481 to 0.61), attributed to minimal semantic redundancy and tokenization constraints. Conversely, long documents showed more moderate improvements (MRR@3 from 0.423 to 0.475), primarily due to embedding truncation that causes loss of context and increased semantic complexity. Future research should focus explicitly on developing hierarchical or segment-based embedding methods to address these limitations.

\paragraph{Ablation Studies}
To clearly understand the impact of the individual components of the framework, we conducted systematic ablation studies (Table~\ref{tab:proposed_strategies}). Training with positive documents alone produced only slight gains (+0.03 MRR@3), reaffirming the critical role of high-quality hard negatives. Evaluating individual embedding models separately indicated varying performance due to their differing semantic representations and underlying training. However, the concatenation of diverse embeddings provided significant performance improvements (+0.15 MRR@3), clearly highlighting the advantages of capturing semantic diversity.

Additionally, PCA-based dimensionality reduction analysis identified the optimal variance threshold at 95\%. Lower thresholds resulted in marked semantic degradation, reducing retrieval performance. This trade-off highlights PCA as an essential efficiency-enhancing step for the framework.

Collectively, these detailed analyses underscore our method's strengths, limitations, and methodological rationale, providing clear empirical justification for each design decision.

\begin{table}[h!]
\centering
% \scalebox{0.85}{ % Adjusted scale factor to ensure the table fits in a single column
\scalebox{0.75}{ % Adjusted scale factor to ensure the table fits in a single column
\begin{tabular}{|c|l|c|c|}
\hline
\textbf{\#} & \textbf{Proposed Strategies}           & \textbf{MRR@3} & \textbf{MRR@10} \\ \hline
1              & Baseline                               & 0.42           & 0.45            \\ \hline
\multicolumn{4}{|c|}{\textbf{Positive Document (PD) Only}}                        \\ \hline
2              & Fine-tuning with PD Only & 0.45           & 0.51            \\ \hline
\multicolumn{4}{|c|}{\textbf{Hard Negative(HN) with Embedding $E_k$}}                         \\ \hline
3a             & HN with $E_1$ + PD                          & 0.45           & 0.51            \\
3b             & HN with $E_2$ + PD                           & 0.47           & 0.53            \\
3c             & HN with $E_3$ + PD                           & 0.51           & 0.55            \\
3d             & HN with $E_4$ + PD                           & 0.45           & 0.52            \\
3e             & HN with $E_5$ + PD                           & 0.48           & 0.51            \\
3f             & HN with $E_6$ + PD                           & 0.49           & 0.52            \\
3g             & HN with X\textsubscript{concat} + PD       & \textbf{0.57}  & \textbf{0.64}   \\ \hline
\multicolumn{4}{|c|}{\textbf{X\textsubscript{PCA} Variance Impact}  + PD }                           \\ \hline
4a             & HN with X\textsubscript{PCA} (99\% Variance) & \uline{0.57}  & \uline{0.64}   \\
4b             & HN with X\textsubscript{PCA} (95\% Variance) & \textbf{0.57} & \textbf{0.64}  \\
4c             & HN with X\textsubscript{PCA} (90\% Variance) & 0.55           & 0.63           \\
4d             & HN with X\textsubscript{PCA} (80\% Variance) & 0.51           & 0.58           \\
4e             & HN with X\textsubscript{PCA} (70\% Variance) & 0.49           & 0.56           \\ \hline
\end{tabular}
}
\caption{Results of ablation study showing the impact of embeddings, PCA variance thresholds, and positive documents on MRR, on the in-house re-ranker model.}
\vspace{-1em}
\label{tab:proposed_strategies}
\end{table}

\subsection{Case Studies: Examples of Hard Negative Impact}
\label{case_study}

\label{case_study}
\begin{figure}[h]
    \centering
    \includegraphics[width=\columnwidth]{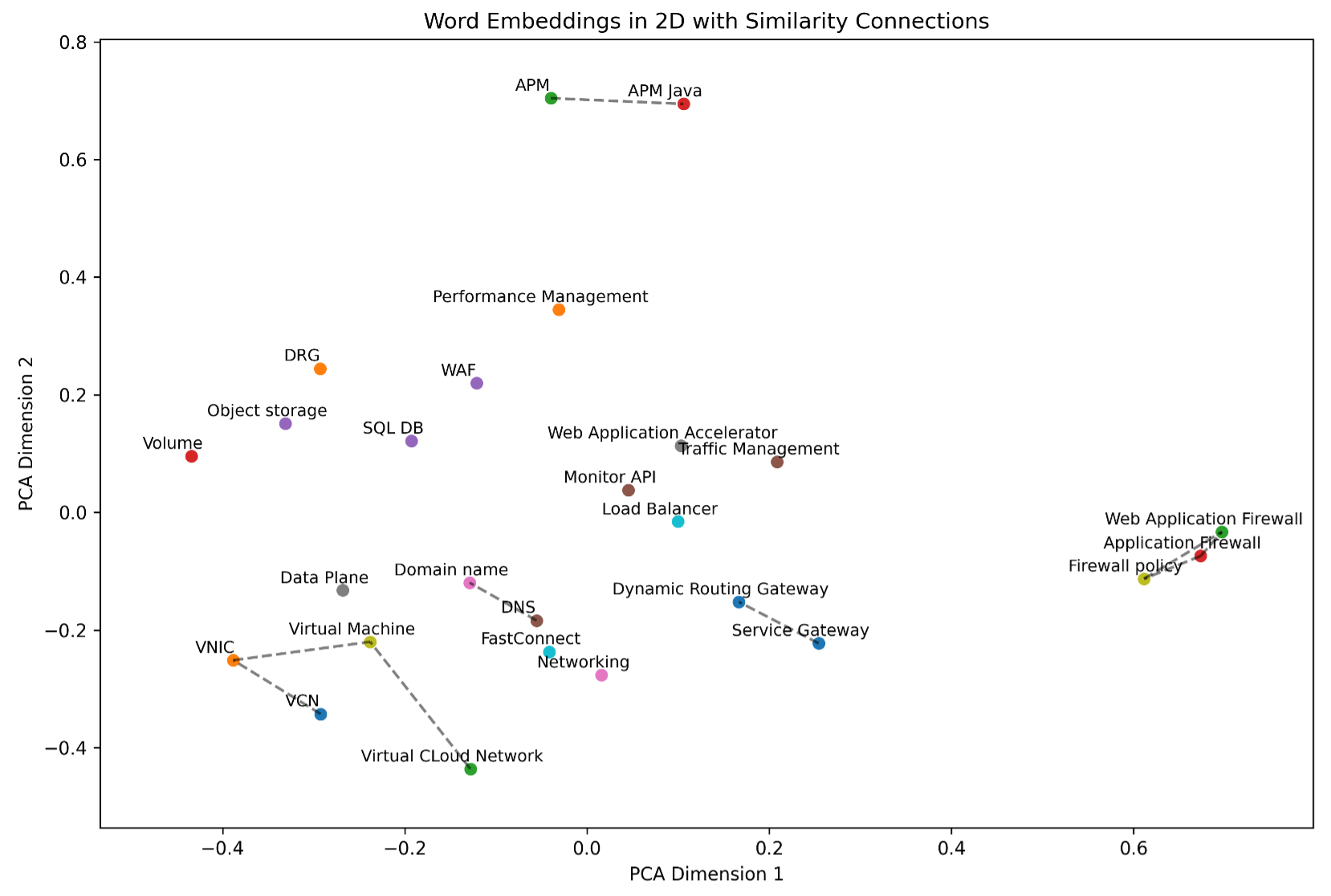} % Placeholder for pipeline diagram
    \caption{Illustrations of similar topics in the domain of Cloud Computing}
    \label{fig:similar_topics}
    \vspace{-1em}
\end{figure}

Figure \ref{fig:similar_topics} shows how similar topics in the domain of cloud computing. To demonstrate the qualitative benefits of the proposed framework, we present two case studies where the baseline and fine-tuned models produce different ranking results. These examples highlight the significance of hard negatives in distinguishing semantically similar but contextually irrelevant documents.

%\begin{figure*}[ht!]

\paragraph{Case Study 1: Disambiguating Technical Acronyms.}
\begin{itemize}
    \item \textbf{Query (Q):} "What is VCN in Cloud Infrastructure?"
    \item \textbf{Positive Document (PD):} A document explaining "Virtual Cloud Network (VCN)" in Cloud Infrastructure, detailing its setup and usage.
    \item \textbf{Hard Negative (HN):} A document discussing "Virtual Network Interface Card (VNIC)" in the context of networking hardware.
\end{itemize}
\textbf{Baseline Result:} The baseline model incorrectly ranks the hard negative above the positive document due to overlapping terms such as "virtual" and "network." \\
\textbf{Proposed Method Result:} The fine-tuned model ranks the positive document higher, correctly identifying the contextual match between the query and the description of VCN. This improvement is attributed to the triplet loss training with hard negatives.

\paragraph{Case Study 2: Domain-Specific Terminology.}
\begin{itemize}
    \item \textbf{Query (Q):} "How does the CI WAF handle incoming traffic?"
    \item \textbf{Positive Document (PD):} A document explaining the Web Application Firewall (WAF) in CI, its configuration, and traffic filtering mechanisms.
    \item \textbf{Hard Negative (HN):} A document discussing general firewall configurations in networking.
\end{itemize}
\textbf{Baseline Result:} The baseline model ranks the hard negative higher due to lexical overlap between the terms "firewall" and "traffic." \\
\textbf{Proposed Method Result:} The proposed framework ranks the positive document higher, leveraging domain-specific semantic representations.

These case studies illustrate the practical advantages of training with hard negatives, especially in domains with overlapping terminology or acronyms.

Additional detailed analyses, illustrative practical implications for enterprise applications, and explicit future directions are discussed in detail in \ref{appendix:results},  and \ref{enterprise_application}.%, and \ref{appendix:limitations}.

% Additional detailed analyses, illustrative case studies, practical implications for enterprise applications, limitations, and explicit future directions are discussed in detail in \ref{appendix:results}, \ref{case_study}, and \ref{enterprise_application}., and \ref{appendix:limitations}.

\section{Conclusion}

We introduced a scalable, modular framework leveraging dynamic ensemble-based hard-negative mining to significantly enhance re-ranking models in enterprise and domain-specific retrieval scenarios. Our method dynamically curates semantically challenging yet contextually irrelevant negatives, allowing re-ranking models to effectively discriminate subtle semantic differences. Empirical evaluations on proprietary enterprise data and diverse public domain-specific benchmarks demonstrated substantial improvements of up to 15\% in MRR@3 and 19\% in MRR@10 over state-of-the-art negative sampling techniques, including BM25, In-Batch Negatives, STAR, and ADORE+STAR.

Our approach offers clear practical benefits in real-world deployments, benefiting  downstream applications such as knowledge management, customer support systems, and Retrieval-Augmented Generation (RAG), where retrieval precision directly influences user satisfaction and Generative AI effectiveness. The strong performance and generalizability across various domains further underscore the framework's readiness for industry-scale deployment.

Future work will focus on extending our framework to handle incremental updates of enterprise knowledge bases and exploring real-time negative sampling strategies for continuously evolving corpora, further enhancing the adaptability and robustness required in practical industry settings.

\section{Limitations}

While our approach advances the state of hard negative mining and encoder-based retrieval, several limitations remain that open avenues for future research. One key challenge is the performance disparity between short and long documents. Addressing this requires more effective document chunking strategies and the development of hierarchical representations to preserve context across segments. Additionally, the retrieval of long documents is complicated by semantic redundancy and truncation, warranting deeper analysis of their structural complexity. Our current use of embedding concatenation for ensembling could also be refined—future work should evaluate alternative fusion techniques such as weighted averaging or attention-based mechanisms. Moreover, extending the retrieval framework to support cross-lingual and multilingual scenarios would enhance its utility in globally distributed applications.

% Bibliography entries for the entire Anthology, followed by custom entries
%\bibliography{anthology,custom}
% Custom bibliography entries only
\bibliography{custom}

\begin{thebibliography}{61}
\providecommand{\natexlab}[1]{#1}

\bibitem[{AGARWAL(2021)}]{Agarwal2021}
AMIT AGARWAL. 2021.
\newblock \href {https://doi.org/10.13140/RG.2.2.33887.53928} {Evaluate generalisation \& robustness of visual features from images to video}.
\newblock \emph{ResearchGate}.
\newblock Available at \url{https://doi.org/10.13140/RG.2.2.33887.53928}.

\bibitem[{Agarwal et~al.(2024{\natexlab{a}})Agarwal, Panda, and Pachauri}]{Agarwal2024}
Amit Agarwal, Srikant Panda, and Kulbhushan Pachauri. 2024{\natexlab{a}}.
\newblock Synthetic document generation pipeline for training artificial intelligence models.
\newblock US Patent App. 17/994,712.

\bibitem[{Agarwal et~al.(2025)Agarwal, Panda, and Pachauri}]{agarwal-etal-2025-fs}
Amit Agarwal, Srikant Panda, and Kulbhushan Pachauri. 2025.
\newblock \href {https://aclanthology.org/2025.coling-industry.9/} {{FS}-{DAG}: Few shot domain adapting graph networks for visually rich document understanding}.
\newblock In \emph{Proceedings of the 31st International Conference on Computational Linguistics: Industry Track}, pages 100--114, Abu Dhabi, UAE. Association for Computational Linguistics.

\bibitem[{Agarwal et~al.(2024{\natexlab{b}})Agarwal, Patel, Pattnayak, Panda, Kumar, and Kumar}]{agarwal2024enhancing}
Amit Agarwal, Hitesh Patel, Priyaranjan Pattnayak, Srikant Panda, Bhargava Kumar, and Tejaswini Kumar. 2024{\natexlab{b}}.
\newblock Enhancing document ai data generation through graph-based synthetic layouts.
\newblock \emph{arXiv preprint arXiv:2412.03590}.

\bibitem[{AI(2023)}]{jina_reranker}
Jina AI. 2023.
\newblock \href {https://huggingface.co/jinaai/jina-reranker-v2-base-multilingual} {jina-reranker-v2-base-multilingual}.

\bibitem[{Askari et~al.(2023)Askari, Aliannejadi, Kanoulas, and Verberne}]{Askari2023}
Arian Askari, Mohammad Aliannejadi, Evangelos Kanoulas, and Suzan Verberne. 2023.
\newblock \href {http://arxiv.org/abs/2305.02320} {Generating synthetic documents for cross-encoder re-rankers: A comparative study of chatgpt and human experts}.

\bibitem[{Bai et~al.(2023)Bai, Guo, Liu, Yang, Liang, Yan, and Li}]{bai2023griprankbridginggapretrieval}
Jiaqi Bai, Hongcheng Guo, Jiaheng Liu, Jian Yang, Xinnian Liang, Zhao Yan, and Zhoujun Li. 2023.
\newblock \href {https://arxiv.org/abs/2305.18144} {Griprank: Bridging the gap between retrieval and generation via the generative knowledge improved passage ranking}.
\newblock \emph{Preprint}, arXiv:2305.18144.

\bibitem[{Castelli et~al.(2019)Castelli, Chakravarti, Dana, Ferritto, Florian, Franz, Garg, Khandelwal, McCarley, McCawley, Nasr, Pan, Pendus, Pitrelli, Pujar, Roukos, Sakrajda, Sil, Uceda-Sosa, Ward, and Zhang}]{castelli2019techqadataset}
Vittorio Castelli, Rishav Chakravarti, Saswati Dana, Anthony Ferritto, Radu Florian, Martin Franz, Dinesh Garg, Dinesh Khandelwal, Scott McCarley, Mike McCawley, Mohamed Nasr, Lin Pan, Cezar Pendus, John Pitrelli, Saurabh Pujar, Salim Roukos, Andrzej Sakrajda, Avirup Sil, Rosario Uceda-Sosa, Todd Ward, and Rong Zhang. 2019.
\newblock \href {https://arxiv.org/abs/1911.02984} {The techqa dataset}.
\newblock \emph{Preprint}, arXiv:1911.02984.

\bibitem[{Cohere(2023{\natexlab{a}})}]{cohere_embed_2023}
Cohere. 2023{\natexlab{a}}.
\newblock Cohere-embed-multilingual-v3.0.
\newblock Available at: \url{https://cohere.com/blog/introducing-embed-v3}.

\bibitem[{Cohere(2023{\natexlab{b}})}]{cohere_reranker_2023}
Cohere. 2023{\natexlab{b}}.
\newblock Reranker model.
\newblock Available at: \url{https://docs.cohere.com/v2/docs/reranking-with-cohere}.

\bibitem[{de~Souza P.~Moreira et~al.(2024)de~Souza P.~Moreira, Osmulski, Xu, Ak, Schifferer, and Oldridge}]{moreira2024nvretrieverimprovingtextembedding}
Gabriel de~Souza P.~Moreira, Radek Osmulski, Mengyao Xu, Ronay Ak, Benedikt Schifferer, and Even Oldridge. 2024.
\newblock \href {https://arxiv.org/abs/2407.15831} {Nv-retriever: Improving text embedding models with effective hard-negative mining}.
\newblock \emph{Preprint}, arXiv:2407.15831.

\bibitem[{Diggelmann et~al.(2021)Diggelmann, Boyd-Graber, Bulian, Ciaramita, and Leippold}]{diggelmann2021climatefeverdatasetverificationrealworld}
Thomas Diggelmann, Jordan Boyd-Graber, Jannis Bulian, Massimiliano Ciaramita, and Markus Leippold. 2021.
\newblock \href {https://arxiv.org/abs/2012.00614} {Climate-fever: A dataset for verification of real-world climate claims}.
\newblock \emph{Preprint}, arXiv:2012.00614.

\bibitem[{Dua et~al.(2024)Dua, Pabolu, and Guo}]{dua2024generating}
Karan Dua, Praneet Pabolu, and Mengqing Guo. 2024.
\newblock Generating templates for use in synthetic document generation processes.
\newblock US Patent App. 18/295,765.

\bibitem[{Dua et~al.(2025)Dua, Pabolu, and Gupta}]{dua2025generation}
Karan Dua, Praneet Pabolu, and Ranjeet~Kumar Gupta. 2025.
\newblock Generation of synthetic doctor-patient conversations.
\newblock US Patent App. 18/495,966.

\bibitem[{Feng et~al.(2022)Feng, Yang, Cer, Arivazhagan, and Wang}]{feng2022languageagnosticbertsentenceembedding}
Fangxiaoyu Feng, Yinfei Yang, Daniel Cer, Naveen Arivazhagan, and Wei Wang. 2022.
\newblock \href {https://arxiv.org/abs/2007.01852} {Language-agnostic bert sentence embedding}.
\newblock \emph{Preprint}, arXiv:2007.01852.

\bibitem[{Gao and Callan(2021)}]{Gao2021}
Luyu Gao and Jamie Callan. 2021.
\newblock \href {https://doi.org/10.18653/V1/2021.EMNLP-MAIN.75} {Condenser: a pre-training architecture for dense retrieval}.
\newblock \emph{EMNLP 2021 - 2021 Conference on Empirical Methods in Natural Language Processing, Proceedings}, pages 981--993.

\bibitem[{Glass et~al.(2022)Glass, Rossiello, Chowdhury, Naik, Cai, and Gliozzo}]{glass-etal-2022-re2g}
Michael Glass, Gaetano Rossiello, Md~Faisal~Mahbub Chowdhury, Ankita Naik, Pengshan Cai, and Alfio Gliozzo. 2022.
\newblock \href {https://doi.org/10.18653/v1/2022.naacl-main.194} {{R}e2{G}: Retrieve, rerank, generate}.
\newblock In \emph{Proceedings of the 2022 Conference of the North American Chapter of the Association for Computational Linguistics: Human Language Technologies}, pages 2701--2715, Seattle, United States. Association for Computational Linguistics.

\bibitem[{Guo et~al.(2023)Guo, Zhang, Wang, Jiang, Nie, Ding, Yue, and Wu}]{Guo2023}
Biyang Guo, Xin Zhang, Ziyuan Wang, Minqi Jiang, Jinran Nie, Yuxuan Ding, Jianwei Yue, and Yupeng Wu. 2023.
\newblock \href {https://arxiv.org/abs/2301.07597v1} {How close is chatgpt to human experts? comparison corpus, evaluation, and detection}.

\bibitem[{Guu et~al.(2020)Guu, Lee, Tung, Pasupat, and Chang}]{Guu2020}
Kelvin Guu, Kenton Lee, Zora Tung, Panupong Pasupat, and Ming-Wei Chang. 2020.
\newblock Realm: Retrieval-augmented language model pre-training.

\bibitem[{Jasila et~al.(2023)Jasila, Saleena, and Abdul~Nazeer}]{jasila2023efficient}
EK~Jasila, N~Saleena, and KA~Abdul~Nazeer. 2023.
\newblock An efficient document clustering approach for devising semantic clusters.
\newblock \emph{Cybernetics and Systems}, pages 1--18.

\bibitem[{Karpukhin et~al.(2020)Karpukhin, Oğuz, Min, Lewis, Wu, Edunov, Chen, and Yih}]{Karpukhin2020}
Vladimir Karpukhin, Barlas Oğuz, Sewon Min, Patrick Lewis, Ledell Wu, Sergey Edunov, Danqi Chen, and Wen~Tau Yih. 2020.
\newblock \href {https://doi.org/10.18653/V1/2020.EMNLP-MAIN.550} {Dense passage retrieval for open-domain question answering}.
\newblock \emph{EMNLP 2020 - 2020 Conference on Empirical Methods in Natural Language Processing, Proceedings of the Conference}, pages 6769--6781.

\bibitem[{Lee et~al.(2024)Lee, Roy, Xu, Raiman, Shoeybi, Catanzaro, and Ping}]{lee2024nv}
Chankyu Lee, Rajarshi Roy, Mengyao Xu, Jonathan Raiman, Mohammad Shoeybi, Bryan Catanzaro, and Wei Ping. 2024.
\newblock Nv-embed: Improved techniques for training llms as generalist embedding models.
\newblock \emph{arXiv preprint arXiv:2405.17428}.

\bibitem[{Li et~al.(2024)Li, Xie, and Zhou}]{10446388}
Fulu Li, Zhiwen Xie, and Guangyou Zhou. 2024.
\newblock \href {https://doi.org/10.1109/ICASSP48485.2024.10446388} {Theme-enhanced hard negative sample mining for open-domain question answering}.
\newblock In \emph{ICASSP 2024 - 2024 IEEE International Conference on Acoustics, Speech and Signal Processing (ICASSP)}, pages 12436--12440.

\bibitem[{Li and Li(2023)}]{li2023angle}
Xianming Li and Jing Li. 2023.
\newblock Angle-optimized text embeddings.
\newblock \emph{arXiv preprint arXiv:2309.12871}.

\bibitem[{Liu et~al.(2021)Liu, Hashimoto, Zhou, Yavuz, Xiong, and Yu}]{Liu2021DenseHR}
Ye~Liu, Kazuma Hashimoto, Yingbo Zhou, Semih Yavuz, Caiming Xiong, and Philip~S. Yu. 2021.
\newblock \href {https://api.semanticscholar.org/CorpusID:240288895} {Dense hierarchical retrieval for open-domain question answering}.
\newblock In \emph{Conference on Empirical Methods in Natural Language Processing}.

\bibitem[{MacAvaney et~al.(2019)MacAvaney, Yates, Cohan, and Goharian}]{MacAvaney2019}
Sean MacAvaney, Andrew Yates, Arman Cohan, and Nazli Goharian. 2019.
\newblock \href {https://doi.org/10.1145/3331184.3331317} {Cedr: Contextualized embeddings for document ranking}.
\newblock \emph{SIGIR 2019 - Proceedings of the 42nd International ACM SIGIR Conference on Research and Development in Information Retrieval}, pages 1101--1104.

\bibitem[{Ma{\'c}kiewicz and Ratajczak(1993)}]{mackiewicz1993principal}
Andrzej Ma{\'c}kiewicz and Waldemar Ratajczak. 1993.
\newblock Principal components analysis (pca).
\newblock \emph{Computers \& Geosciences}, 19(3):303--342.

\bibitem[{McInnes et~al.(2020)McInnes, Healy, and Melville}]{mcinnes2020umapuniformmanifoldapproximation}
Leland McInnes, John Healy, and James Melville. 2020.
\newblock \href {https://arxiv.org/abs/1802.03426} {Umap: Uniform manifold approximation and projection for dimension reduction}.
\newblock \emph{Preprint}, arXiv:1802.03426.

\bibitem[{Meghwani(2024)}]{meghwani2024enhancingretrievalperformanceensemble}
Hansa Meghwani. 2024.
\newblock \href {https://arxiv.org/abs/2411.02404} {Enhancing retrieval performance: An ensemble approach for hard negative mining}.
\newblock \emph{Preprint}, arXiv:2411.02404.

\bibitem[{Mehta et~al.(2024)Mehta, Agarwal, and Kaliyar}]{mehta2024comprehensive}
Vivek Mehta, Mohit Agarwal, and Rohit~Kumar Kaliyar. 2024.
\newblock A comprehensive and analytical review of text clustering techniques.
\newblock \emph{International Journal of Data Science and Analytics}, pages 1--20.

\bibitem[{Meng et~al.(2024{\natexlab{a}})Meng, Liu, Joty, Xiong, Zhou, and Yavuz}]{SFR-embedding-2}
Rui Meng, Ye~Liu, Shafiq~Rayhan Joty, Caiming Xiong, Yingbo Zhou, and Semih Yavuz. 2024{\natexlab{a}}.
\newblock \href {https://huggingface.co/Salesforce/SFR-Embedding-2_R} {Sfr-embedding-2: Advanced text embedding with multi-stage training}.

\bibitem[{Meng et~al.(2024{\natexlab{b}})Meng, Liu, Joty, Xiong, Zhou, and Yavuz}]{SFRAIResearch2024}
Rui Meng, Ye~Liu, Shafiq~Rayhan Joty, Caiming Xiong, Yingbo Zhou, and Semih Yavuz. 2024{\natexlab{b}}.
\newblock \href {https://blog.salesforceairesearch.com/sfr-embedded-mistral/} {Sfr-embedding-mistral: Enhance text retrieval with transfer learning}.
\newblock Salesforce AI Research Blog.

\bibitem[{Nguyen et~al.(2022)Nguyen, Bui, Vuong, and Phan}]{Nguyen2022}
Thanh-Do Nguyen, Chi~Minh Bui, Thi-Hai-Yen Vuong, and Xuan-Hieu Phan. 2022.
\newblock Passage-based bm25 hard negatives: A simple and effective negative sampling strategy for dense retrieval.

\bibitem[{Nogueira and Cho(2019)}]{Nogueira2019passage}
Rodrigo Nogueira and Kyunghyun Cho. 2019.
\newblock \href {https://arxiv.org/abs/1901.04085v5} {Passage re-ranking with bert}.

\bibitem[{Nogueira et~al.(2019)Nogueira, Yang, Cho, and Lin}]{Nogueira2019}
Rodrigo Nogueira, Wei Yang, Kyunghyun Cho, and Jimmy Lin. 2019.
\newblock Multi-stage document ranking with bert.

\bibitem[{Nussbaum et~al.(2024)Nussbaum, Morris, Duderstadt, and Mulyar}]{nussbaum2024nomic}
Zach Nussbaum, John~X. Morris, Brandon Duderstadt, and Andriy Mulyar. 2024.
\newblock \href {https://arxiv.org/abs/2402.01613} {Nomic embed: Training a reproducible long context text embedder}.
\newblock \emph{Preprint}, arXiv:2402.01613.

\bibitem[{Pabolu et~al.(2024{\natexlab{a}})Pabolu, Dua, and Chaudhury}]{pabolu2024multi1}
Praneet Pabolu, Karan Dua, and Sriram Chaudhury. 2024{\natexlab{a}}.
\newblock Multi-lingual natural language generation.
\newblock US Patent App. 18/318,315.

\bibitem[{Pabolu et~al.(2024{\natexlab{b}})Pabolu, Dua, and Chaudhury}]{pabolu2024multi}
Praneet Pabolu, Karan Dua, and Sriram Chaudhury. 2024{\natexlab{b}}.
\newblock Multi-lingual natural language generation.
\newblock US Patent App. 18/318,327.

\bibitem[{Panda et~al.(2025{\natexlab{a}})Panda, Agarwal, Nambirajan, and Pachauri}]{panda2025out}
Srikant Panda, Amit Agarwal, Gouttham Nambirajan, and Kulbhushan Pachauri. 2025{\natexlab{a}}.
\newblock Out of distribution element detection for information extraction.
\newblock US Patent App. 18/347,983.

\bibitem[{Panda et~al.(2025{\natexlab{b}})Panda, Agarwal, and Pachauri}]{panda2025techniques}
Srikant Panda, Amit Agarwal, and Kulbhushan Pachauri. 2025{\natexlab{b}}.
\newblock Techniques of information extraction for selection marks.
\newblock US Patent App. 18/240,344.

\bibitem[{Patel et~al.(2025)Patel, Agarwal, Das, Kumar, Panda, Pattnayak, Rafi, Kumar, and Chae}]{patel2025sweeval}
Hitesh~Laxmichand Patel, Amit Agarwal, Arion Das, Bhargava Kumar, Srikant Panda, Priyaranjan Pattnayak, Taki~Hasan Rafi, Tejaswini Kumar, and Dong-Kyu Chae. 2025.
\newblock Sweeval: Do llms really swear? a safety benchmark for testing limits for enterprise use.
\newblock In \emph{Proceedings of the 2025 Conference of the Nations of the Americas Chapter of the Association for Computational Linguistics: Human Language Technologies (Volume 3: Industry Track)}, pages 558--582.

\bibitem[{Patel et~al.(2024)Patel, Agarwal, Kumar, Gupta, and Pattnayak}]{patel2024llm}
Hitesh~Laxmichand Patel, Amit Agarwal, Bhargava Kumar, Karan Gupta, and Priyaranjan Pattnayak. 2024.
\newblock Llm for barcodes: Generating diverse synthetic data for identity documents.
\newblock \emph{arXiv preprint arXiv:2411.14962}.

\bibitem[{Pattnayak et~al.(2025{\natexlab{a}})Pattnayak, Agarwal, Meghwani, Patel, and Panda}]{pattnayak2025hybrid}
Priyaranjan Pattnayak, Amit Agarwal, Hansa Meghwani, Hitesh~Laxmichand Patel, and Srikant Panda. 2025{\natexlab{a}}.
\newblock Hybrid ai for responsive multi-turn online conversations with novel dynamic routing and feedback adaptation.
\newblock In \emph{Proceedings of the 4th International Workshop on Knowledge-Augmented Methods for Natural Language Processing}, pages 215--229.

\bibitem[{Pattnayak et~al.(2025{\natexlab{b}})Pattnayak, Patel, and Agarwal}]{pattnayak2025tokenizationmattersimprovingzeroshot}
Priyaranjan Pattnayak, Hitesh~Laxmichand Patel, and Amit Agarwal. 2025{\natexlab{b}}.
\newblock \href {https://arxiv.org/abs/2504.16977} {Tokenization matters: Improving zero-shot ner for indic languages}.
\newblock \emph{Preprint}, arXiv:2504.16977.

\bibitem[{Pattnayak et~al.(2025{\natexlab{c}})Pattnayak, Patel, Agarwal, Kumar, Panda, and Kumar}]{pattnayak2025clinicalqa20multitask}
Priyaranjan Pattnayak, Hitesh~Laxmichand Patel, Amit Agarwal, Bhargava Kumar, Srikant Panda, and Tejaswini Kumar. 2025{\natexlab{c}}.
\newblock \href {https://arxiv.org/abs/2502.13108} {Clinical qa 2.0: Multi-task learning for answer extraction and categorization}.
\newblock \emph{Preprint}, arXiv:2502.13108.

\bibitem[{Pradeep et~al.(2022)Pradeep, Liu, Zhang, Li, Yates, and Lin}]{Pradeep2022}
Ronak Pradeep, Yuqi Liu, Xinyu Zhang, Yilin Li, Andrew Yates, and Jimmy Lin. 2022.
\newblock \href {https://doi.org/10.1007/978-3-030-99736-6_44} {Squeezing water from a stone: A bag of tricks for further improving cross-encoder effectiveness for reranking}.
\newblock In \emph{Lecture Notes in Computer Science (including subseries Lecture Notes in Artificial Intelligence and Lecture Notes in Bioinformatics)}, volume 13185 LNCS, pages 655--670. Springer Science and Business Media Deutschland GmbH.

\bibitem[{Qu et~al.(2020)Qu, Ding, Liu, Liu, Ren, Zhao, Dong, Wu, and Wang}]{Qu2020}
Yingqi Qu, Yuchen Ding, Jing Liu, Kai Liu, Ruiyang Ren, Wayne~Xin Zhao, Daxiang Dong, Hua Wu, and Haifeng Wang. 2020.
\newblock Rocketqa: An optimized training approach to dense passage retrieval for open-domain question answering.

\bibitem[{Raffel et~al.(2020)Raffel, Shazeer, Roberts, Lee, Narang, Matena, Zhou, Li, and Liu}]{2020t5}
Colin Raffel, Noam Shazeer, Adam Roberts, Katherine Lee, Sharan Narang, Michael Matena, Yanqi Zhou, Wei Li, and Peter~J. Liu. 2020.
\newblock \href {http://jmlr.org/papers/v21/20-074.html} {Exploring the limits of transfer learning with a unified text-to-text transformer}.
\newblock \emph{Journal of Machine Learning Research}, 21(140):1--67.

\bibitem[{Reimers and Gurevych(2019)}]{reimers-2019-sentence-bert}
Nils Reimers and Iryna Gurevych. 2019.
\newblock \href {https://huggingface.co/sentence-transformers/all-mpnet-base-v2} {Sentence-bert: Sentence embeddings using siamese bert-networks}.
\newblock In \emph{Proceedings of the 2019 Conference on Empirical Methods in Natural Language Processing}.

\bibitem[{Robertson and Walker(1994)}]{Robertson1994}
S.~E. Robertson and S.~Walker. 1994.
\newblock \href {https://doi.org/10.1007/978-1-4471-2099-5_24} {\emph{Some Simple Effective Approximations to the 2-Poisson Model for Probabilistic Weighted Retrieval}}, pages 232--241.
\newblock Springer London.

\bibitem[{Sturua et~al.(2024)Sturua, Mohr, Akram, Günther, Wang, Krimmel, Wang, Mastrapas, Koukounas, Koukounas, Wang, and Xiao}]{sturua2024jinaembeddingsv3multilingualembeddingstask}
Saba Sturua, Isabelle Mohr, Mohammad~Kalim Akram, Michael Günther, Bo~Wang, Markus Krimmel, Feng Wang, Georgios Mastrapas, Andreas Koukounas, Andreas Koukounas, Nan Wang, and Han Xiao. 2024.
\newblock \href {https://arxiv.org/abs/2409.10173} {jina-embeddings-v3: Multilingual embeddings with task lora}.
\newblock \emph{Preprint}, arXiv:2409.10173.

\bibitem[{TheFinAI(2018)}]{fiqa_dataset}
TheFinAI. 2018.
\newblock \href {https://huggingface.co/datasets/TheFinAI/fiqa-sentiment-classification} {Fiqa: A financial question answering dataset}.
\newblock Available at Hugging Face.

\bibitem[{Van~der Maaten and Hinton(2008)}]{van2008visualizing}
Laurens Van~der Maaten and Geoffrey Hinton. 2008.
\newblock Visualizing data using t-sne.
\newblock \emph{Journal of machine learning research}, 9(11).

\bibitem[{Wang et~al.(2023)Wang, Yang, Huang, Yang, Majumder, and Wei}]{wang2023improving}
Liang Wang, Nan Yang, Xiaolong Huang, Linjun Yang, Rangan Majumder, and Furu Wei. 2023.
\newblock Improving text embeddings with large language models.
\newblock \emph{arXiv preprint arXiv:2401.00368}.

\bibitem[{Wold et~al.(1987)Wold, Esbensen, Esbensen, Geladi, and Geladi}]{svante_wold__1987}
Svante Wold, Kim~H. Esbensen, Kim~H. Esbensen, Paul Geladi, and Paul Geladi. 1987.
\newblock \href {https://doi.org/10.1016/0169-7439(87)80084-9} {Principal component analysis}.
\newblock \emph{Chemometrics and Intelligent Laboratory Systems}, 2:37--52.

\bibitem[{Xiao et~al.(2023)Xiao, Liu, Zhang, and Muennighoff}]{bge_embedding}
Shitao Xiao, Zheng Liu, Peitian Zhang, and Niklas Muennighoff. 2023.
\newblock \href {https://arxiv.org/abs/2309.07597} {C-pack: Packaged resources to advance general chinese embedding}.
\newblock \emph{Preprint}, arXiv:2309.07597.

\bibitem[{Xiong et~al.(2020)Xiong, Xiong, Li, Tang, Liu, Bennett, Ahmed, and Overwijk}]{Xiong2020}
Lee Xiong, Chenyan Xiong, Ye~Li, Kwok-Fung Tang, Jialin Liu, Paul~N. Bennett, Junaid Ahmed, and Arnold Overwijk. 2020.
\newblock \href {http://aka.ms/ance.} {Approximate nearest neighbor negative contrastive learning for dense text retrieval}.

\bibitem[{Yang et~al.(2024)Yang, Shao, Dong, and Tang}]{yang2024trisamplerbetternegativesampling}
Zhen Yang, Zhou Shao, Yuxiao Dong, and Jie Tang. 2024.
\newblock \href {https://arxiv.org/abs/2402.11855} {Trisampler: A better negative sampling principle for dense retrieval}.
\newblock \emph{Preprint}, arXiv:2402.11855.

\bibitem[{Zhan et~al.(2021)Zhan, Mao, Liu, Guo, Zhang, and Ma}]{Zhan2021}
Jingtao Zhan, Jiaxin Mao, Yiqun Liu, Jiafeng Guo, Min Zhang, and Shaoping Ma. 2021.
\newblock \href {https://doi.org/10.1145/3404835.3462880} {Optimizing dense retrieval model training with hard negatives}.
\newblock \emph{SIGIR 2021 - Proceedings of the 44th International ACM SIGIR Conference on Research and Development in Information Retrieval}, pages 1503--1512.

\bibitem[{Zhang(2024)}]{dunzhang2024}
Dun Zhang. 2024.
\newblock \href {https://huggingface.co/dunzhang/stella_en_400M_v5} {stella-embedding-model-2024}.

\bibitem[{Zhang et~al.(2024)Zhang, Zhang, Long, Xie, Dai, Tang, Lin, Yang, Xie, Huang et~al.}]{zhang2024mgte}
Xin Zhang, Yanzhao Zhang, Dingkun Long, Wen Xie, Ziqi Dai, Jialong Tang, Huan Lin, Baosong Yang, Pengjun Xie, Fei Huang, et~al. 2024.
\newblock mgte: Generalized long-context text representation and reranking models for multilingual text retrieval.
\newblock \emph{arXiv preprint arXiv:2407.19669}.

\end{thebibliography}

\appendix

\section{Appendix}

\subsection{Extended Related Work}
\label{sec:appendix_related_works}

\paragraph{Hard Negatives in Retrieval Models}
Static and dynamic hard negatives have been extensively studied. Static negatives, such as those generated by BM25~\cite{Robertson1994} or PassageBM25~\cite{Nguyen2022}, provide challenging lexical contrasts but risk overfitting due to their fixed nature~\cite{Qu2020}. Dynamic negatives, as introduced in ANCE~\cite{Xiong2020} and ADORE~\cite{Zhan2021} adapt during training, other effective methods like positive-aware mining ~\cite{moreira2024nvretrieverimprovingtextembedding}, theme-enhanced negatives ~\cite{10446388} offers relevant challenges but incurring high computational costs due to periodic re-indexing and bigger embedding dimension. Our framework mitigates these issues by leveraging clustering and dimensionality reduction to dynamically identify negatives without requiring re-indexing.

Localized Contrastive Estimation (LCE)~\cite{Guo2023,Agarwal2021} further demonstrated the effectiveness of incorporating hard negatives into cross-encoder training, improving reranking accuracy when negatives align with retriever outputs. Additionally, ~\cite{Pradeep2022} highlighted the importance of hard negatives even in advanced pretraining setups like Condenser~\cite{Gao2021}, which emphasizes their necessity for robust optimization.

\paragraph{Advances in Dense Retrieval and Cross-Encoders}
Dense retrieval models like DPR~\cite{Karpukhin2020} and REALM~\cite{Guu2020} encode queries and documents into dense embeddings, enabling semantic matching. Recent advances in dense retrieval and ranking include GripRank's generative knowledge-driven passage ranking ~\cite{bai2023griprankbridginggapretrieval}, Dense Hierarchical Retrieval's multi-stage framework for efficient question answering ~\cite{Liu2021DenseHR,pattnayak2025hybrid,pattnayak2025clinicalqa20multitask,pattnayak2025tokenizationmattersimprovingzeroshot,patel2025sweeval}, and TriSampler's optimized negative sampling for dense retrieval ~\cite{yang2024trisamplerbetternegativesampling}, collectively enhancing retrieval performance.Cross-encoders, such as monoBERT~\cite{Nogueira2019,Nogueira2019passage}, further improve retrieval precision by jointly encoding query-document pairs but require high-quality training data, particularly challenging negatives~\cite{MacAvaney2019,panda2025techniques}. Techniques such as synthetic data generation~\cite{Askari2023, Agarwal2024,agarwal-etal-2025-fs} augment training datasets but lack the realism and semantic depth provided by our hard negative mining approach.

\paragraph{Dimensionality Reduction in IR}
Clustering methods have been used to group semantically similar documents, improving retrieval efficiency and training data organization~\cite{mehta2024comprehensive,jasila2023efficient,dua2025generation,panda2025out}. Dimensionality reduction techniques like PCA~\cite{svante_wold__1987} enhance scalability by reducing computational complexity. Our framework uniquely combines these techniques to dynamically identify negatives that challenge retrieval models in a scalable manner.

\paragraph{Synthetic Data in Retrieval}
Recent work~\cite{Askari2023, Agarwal2024,agarwal2024enhancing,patel2024llm,dua2024generating,pabolu2024multi1,pabolu2024multi} has explored using large language models to generate synthetic training data for retrieval tasks. While effective in low-resource settings, synthetic data often struggles with factual inaccuracies and domain-specific relevance. In contrast, our framework relies on real-world data to curate semantically challenging negatives, ensuring high-quality training samples without introducing synthetic biases.

\paragraph{Summary of Contributions}
While previous works address various aspects of negative sampling, hard negatives, and synthetic data, our approach bridges the gap between static and dynamic strategies. By dynamically curating negatives using clustering and dimensionality reduction, we achieve a scalable and semantically precise methodology tailored to domain-specific retrieval tasks.

\subsection{Extended Methodology}
\label{sec:extended_methodoly}

\subsubsection{Dataset Statistics}
\label{appendix:data_stats}

\begin{figure}[!ht]
    \centering
    \includegraphics[width=1.1\columnwidth]{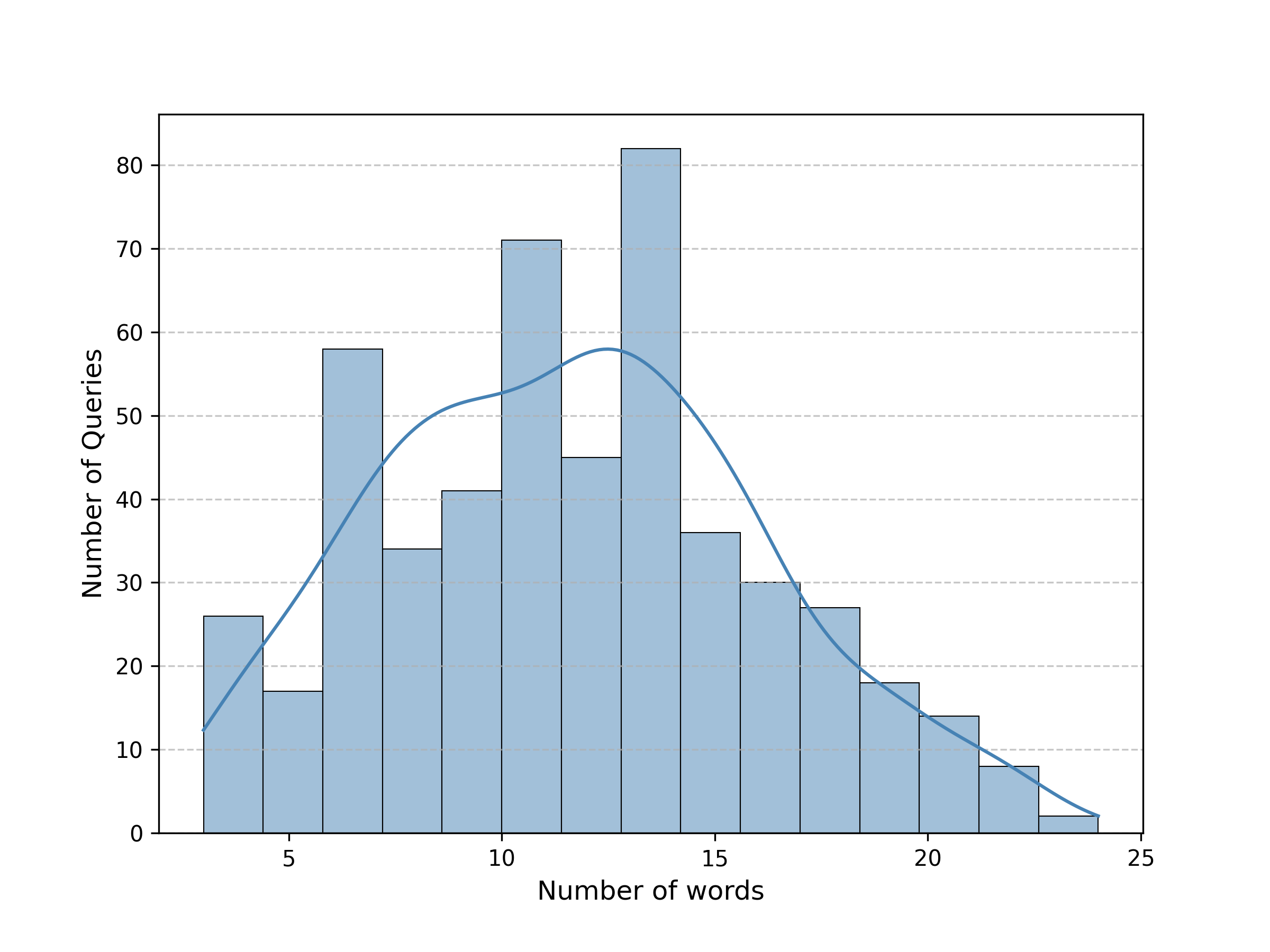} % Replace with your figure file
    \caption{Length Distribution of queries in the dataset.}
    \label{fig:query_length}
\end{figure}

\paragraph{Queries Length Distribution}
In this section we analyze the distribution of queries length in our enterprise dataset. Figure \ref{fig:query_length} shows that the length of queries ranges from 1 to 25 words, with some queries having very few words. This highlights that user queries can sometime be just 2-3 words about a topic, increasing the probability of retrieving documents mentioning those topics or concepts which can be contextually different. Therefore, when we select hard negatives, it is
crucial to consider not only the relationship between the query and documents but also the relationship
between the positive document and other documents, ensuring a comparison with texts on similar topics and similar lengths.

\begin{table}[ht!]
\centering
\scalebox{0.7}{ % Adjust the scale factor as needed
\begin{tabular}{lccc}
\toprule
\textbf{Model ($E_k$)} & \textbf{Params (M)} & \textbf{Dimension} & \textbf{Max Tokens} \\ \midrule
stella\_en\_400M\_v5           & 435   & 8192   & 8192   \\
jina-embeddings-v3             & 572   & 1024   & 8194\\
(multilingual)                    \\
mxbai-embed-large-v1           & 335   & 1024   & 512    \\
bge-large-en-v1.5              & 335   & 1024   & 512    \\
LaBSE                         & 471   & 768    & 256 \\
(multilingual)                     \\
all-mpnet-base-v2              & 110   & 768    & 514  \\
(multilingual)                    \\
% gte-large-en-v1.5              & 434   & 1024   & 8192   \\ 

\bottomrule
\end{tabular}}
\caption{Embedding models used to construct \( X_{\text{concat}} \), combining diverse semantic representations for queries (\( Q \)), positive documents (\( PD \)), and corpus documents (\( D \)).}
\label{tab:embedding_models}
\end{table}

\paragraph{Document Length Distribution}
\begin{figure}[!ht]
    \centering
    \includegraphics[width=1.1\columnwidth]{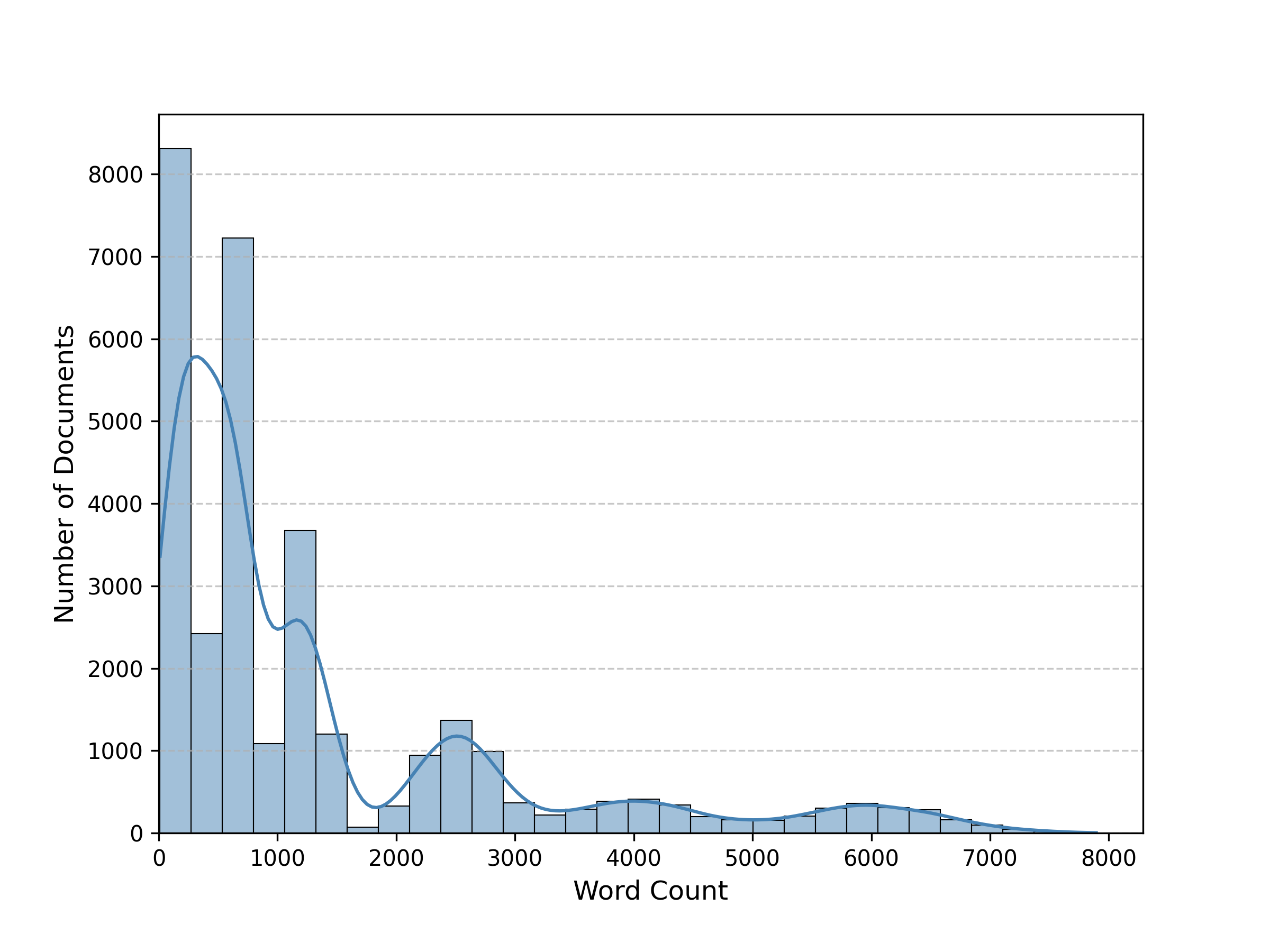} % Replace with your figure file
    \caption{Shows document length distribution in Enterprise corpus. }
    \label{fig:doc_length}
\end{figure}

As shown in Figure \ref{fig:doc_length} , document lengths are significantly longer than query lengths. This disparity
in context length affects the similarity scores, potentially reducing the accuracy of retrieval systems. In our in-house dataset, each query is paired with a single correct document (though its not limited by number of positive-negative document per query). This positive document is crucial for identifying challenging hard negatives and hence helpful for encoder-based model training.

\subsubsection{Embedding Models}
\label{appendix:embedding_models}
Table~\ref{tab:embedding_models} lists the embedding models ~\cite{dunzhang2024,sturua2024jinaembeddingsv3multilingualembeddingstask,li2023angle,bge_embedding,feng2022languageagnosticbertsentenceembedding, reimers-2019-sentence-bert,zhang2024mgte} used to construct \( X_{\text{concat}} \), combining diverse semantic representations for queries (\( Q \)), positive documents (\( PD \)), and corpus documents (\( D \)). These models were selected for their performance, model size, ability to handle  multilingual context,  providing complementary strengths in dimensionality and token coverage. By integrating embeddings from these models, the framework captures nuanced semantic relationships crucial for reranker training.

\subsubsection{Unified Contrastive Loss}
\label{appendix:loss_functions}

The unified contrastive loss is designed to improve ranking precision for both bi-encoders and cross-encoders, by ensuring that positive documents ($PD$) are ranked closer to the query ($Q$) than hard negatives ($D_{HN}$) by a margin $m$. The loss is defined as:
\begin{equation}
L = \sum_{i=1}^{N} \max\left(0, m + d(Q_i, PD_i) - d(Q_i, D_{HN_i})\right)
\end{equation}

where:
\begin{itemize}
    \item $PD_i$: Positive document associated with query $Q_i$.
    \item $D_{HN_i}$: Hard negative document, semantically similar to $PD_i$ but contextually irrelevant.
    \item $d(Q_i, D_i)$: Distance metric measuring relevance between $Q_i$ and $D_i$.
    \item $m$: Margin ensuring $PD_i$ is closer to $Q_i$ than $D_{HN_i}$ by at least $m$, encouraging the model to distinguish between relevant and irrelevant documents effectively.
\end{itemize}
For \textbf{bi-encoders}, the distance metric is defined as:
\begin{equation}
d(Q_i, D_i) = 1 - \text{cosine}(e_{Q_i}, e_{D_i}),
\end{equation}
where $e_{Q_i}$ and $e_{D_i}$ are the embeddings of the query and document, respectively.
For \textbf{cross-encoders}, the distance metric is:
\begin{equation}
d(Q_i, D_i) = -s(Q_i, D_i),
\end{equation}
where $s(Q_i, D_i)$ is the cross-encoder's relevance score for the query-document pair.

This formulation leverages the triplet of ($Q$, $PD$, $D_{HN}$) to minimize $d(Q_i, PD_i)$, pulling positive documents closer to the query, while maximizing $d(Q_i, D_{HN_i})$, pushing hard negatives further away. By emphasizing semantically challenging examples, the model learns sharper decision boundaries for improved ranking precision.

\subsection{Experimental Setup}
\label{appenix:setup}
\paragraph{Datasets} We evaluate our framework extensively using both proprietary and public datasets:
\begin{itemize}
    \item \textbf{Internal Proprietary Dataset:} Consisting of approximately \textit{5250} query-document pairs, on cloud services like computing, networking, firewall, ai services. It includes both short (< \textit{[1024 tokens]}) and long documents (>= \textit{[1024 tokens]}).
    \item \textbf{FiQA Dataset:} A financial domain-specific dataset widely used for retrieval benchmarking.
    \item \textbf{Climate-FEVER Dataset:} An environment-specific fact-checking dataset focused on climate-related information retrieval.
    \item \textbf{TechQA Dataset:} A technical question-answering dataset emphasizing software engineering and technology-related queries.
\end{itemize}

\paragraph{Training and Fine-tuning}
All re-ranking models are fine-tuned using a triplet loss with margin with same hyper-parameters. Early stopping is employed based on validation MRR@10 scores to prevent overfitting.

\paragraph{Evaluation Metrics}
Model performance is evaluated using standard retrieval metrics: Mean Reciprocal Rank (MRR) at positions 3 and 10 (MRR@3 and MRR@10), which measure retrieval quality and ranking precision. Each reported metric is averaged across three experimental runs for robustness.

\subsection{Extended Results \& Ablation}
\label{appendix:results}

\begin{table}[h!]
\centering
\scalebox{0.8}{
\begin{tabular}{l c c c}
\toprule
\textbf{Strategy} & \textbf{\shortstack{Training Data}} & \textbf{MRR@3} & \textbf{MRR@10} \\ \midrule
Baseline          & 0   & 0.42  & 0.45  \\ \hline
\multirow{10}{*}{\shortstack[l]{Finetuned with\\Hard Negatives \\(Ours)}} & 100  & 0.46  & 0.49  \\ 
                   & 200  & 0.48  & 0.51  \\ 
                   & 300  & 0.50  & 0.53  \\ 
                   & 400  & 0.52  & 0.56  \\ 
                   & 500  & 0.52  & 0.58  \\ 
                   & 600  & 0.54  & 0.60  \\ 
                   & 700  & 0.54  & 0.62  \\ 
                   & 800  & 0.56  & 0.63  \\ 
                   & 900  & 0.57  & 0.64  \\ 
                   & 1000 & \textbf{0.57}  & \textbf{0.64}  \\ \bottomrule
\end{tabular}}
\caption{Comparison of Strategies with Varying Training Data Sizes}
\label{tab:strategy_comparison}
\end{table}

\paragraph{Impact of Training Data Size}
\label{appendix:training_datasize}

As shown in Table~\ref{tab:strategy_comparison}, both MRR@3 and MRR@10 improve as the training data size increases, with more pronounced gains in MRR@10. MRR@3 shows gradual improvement, from 0.42 at the baseline to 0.57 with 100 examples, highlighting the model’s enhanced ability to rank relevant documents within the top 3. MRR@10, on the other hand, shows more significant improvement, from 0.45 to 0.64, indicating that the model benefits more from additional data when considering the top 10 ranked documents.

Our method shows promising results even with smaller training sets, demonstrating the effectiveness of incorporating hard negatives early in the training process. This suggests that hard negatives significantly enhance the model's ability to distinguish relevant from irrelevant documents against a given query, even when data is limited. This approach is particularly beneficial in enterprise contexts, where annotated data may be scarce, enabling quicker improvements in domain-specific retrieval performance.

\paragraph{Models in the Study}
In our study we compared the performance of other finetuned re-ranker ~\cite{glass-etal-2022-re2g,wang2023improving, 2020t5} and embedding models ~\cite{zhang2024mgte,nussbaum2024nomic} using hard negatives generated by our proposed framework in Table \ref{tab:open_source_encoders}. We benchmarked the BGE-Reranker~\cite{bge_embedding}, NV-Embed ~\cite{lee2024nv} Salesforce-SFR~\cite{SFR-embedding-2,SFRAIResearch2024} , jina-reranker ~\cite{jina_reranker} and Cohere-Reranker~\cite{cohere_embed_2023,cohere_reranker_2023},  

\subsubsection{Analysis of Long vs. Short Documents}
\label{appendix:long_short_docs}

Table ~\ref{tab:short_long_docs} reveals a consistent disparity in MRR scores between short and long documents, with long documents showing lower performance. Here, we analyze potential reasons and propose mitigation strategies.

\paragraph{Challenges with Long Documents.}
\begin{itemize}
    \item \textbf{Semantic Redundancy:} Long documents often contain repetitive or tangential content, diluting their relevance to a specific query.
    \item \textbf{Context Truncation:} Fixed-length tokenization (e.g., 512 or 1024 tokens) truncates long documents, potentially discarding critical information.
    \item \textbf{Query-to-Document Mismatch:} Short queries may not provide sufficient context to match the nuanced information spread across a lengthy document.
\end{itemize}

\paragraph{Potential Solutions.}
\begin{itemize}
    \item \textbf{Chunk-Based Retrieval:} Split long documents into smaller, semantically coherent chunks and rank them individually.
    \item \textbf{Hierarchical Embeddings:} Use hierarchical models to aggregate sentence- or paragraph-level embeddings for better context representation.
    \item \textbf{Query Expansion:} Enhance short queries with additional context using techniques like query rewriting or pseudo-relevance feedback.
\end{itemize}

This analysis highlights the need for future work to address the inherent challenges of ranking long documents effectively.

\subsection{Practical Implications for Enterprise Applications}
\label{enterprise_application}
The proposed framework has significant practical implications for enterprise information retrieval systems, particularly in retrieval-augmented generation (RAG) pipelines.

\paragraph{Improved Ranking Precision.}
By training with hard negatives, the model ensures that the most relevant documents are retrieved for each query. This is particularly critical for enterprise use cases such as:
\begin{itemize}
    \item \textbf{Technical Support:} Retrieving precise documentation for customer queries, reducing resolution times.
    \item \textbf{Knowledge Management:} Ensuring that employees access the most relevant internal resources quickly.
\end{itemize}

\paragraph{Enhanced Generative Quality.}
High-quality retrieval directly improves the factual accuracy and coherence of outputs generated by large language models in RAG pipelines. For example:
\begin{itemize}
    \item \textbf{Documentation Summarization:} Summaries generated by models like GPT are more reliable when based on top-ranked, accurate sources.
    \item \textbf{Customer Interaction:} Chatbots generate more contextually relevant responses when fed precise retrieved documents.
\end{itemize}

\paragraph{Scalability and Adaptability.}
The framework’s modular design, including the use of diverse embeddings and clustering-based hard negative selection, allows it to adapt to:
\begin{itemize}
    \item Different industries (e.g., healthcare, finance, manufacturing).
    \item Multi-lingual or cross-lingual retrieval tasks.
\end{itemize}

These practical implications underscore the versatility and enterprise readiness of the proposed framework.

\end{document}